\documentclass[11pt,a4paper]{article}
\pdfoutput=1
\usepackage[utf8]{inputenc}
\usepackage{jheppub}
\usepackage{graphicx}
\usepackage{subfig}

\def\ngluon{{\sc NGluon}}
\def\njet{{\sc NJet}}
\def\njettwo{{\sc NJet 2.0}}

\def\qb{\bar{q}}
\def\GeV{{\rm GeV}}
\def\eps{\epsilon}

\preprint{MPP-2013-319, CERN-PH-TH/2013-318}

\title{Next-to-leading order QCD corrections to di-photon production in association with up to three
jets at the Large Hadron Collider}

\author[a]{Simon Badger}
\author[b]{Alberto Guffanti}
\author[c]{Valery Yundin}
\affiliation[a]{
Theory Division, Physics Department, CERN, CH-1211 Geneva 23, Switzerland}
\affiliation[b]{
Niels Bohr International Academy and Discovery Center, The Niels Bohr Institute,\\
University of Copenhagen, Blegdamsvej 17, DK-2100 Copenhagen, Denmark}
\affiliation[c]{
Max-Planck-Institut für Physik, Föhringer Ring 6, 80805 Munich, Germany}

\emailAdd{simon.badger@cern.ch}
\emailAdd{alberto.guffanti@nbi.dk}
\emailAdd{yundin@mpp.mpg.de}

\abstract{
We present the computation of next-to-leading order (NLO) QCD corrections to
di-photon production in association with two or three hard jets in $pp$
collisions at a center-of-mass energy of 8~TeV. The inclusion of NLO
corrections is shown to substantially reduce the theoretical uncertainties
estimated from scale variations on total cross section predictions. We study a
range of differential distributions relevant for phenomenological studies of photon pair
production in association with jets at the LHC. Using an efficient
computational set-up we performed a detailed study of uncertainties due to
parton distribution functions. The computation of the virtual corrections is
performed using new features of the C++ library \njet.
}

\keywords{}

\begin{document}

\maketitle
\flushbottom

\section{Introduction}

There has been a lot of recent interest in the study of $pp\to\gamma\gamma+{}$jets processes as
an important background to $pp\to H \to \gamma\gamma$ which is one of the cleanest decay 
channels for studies of the Higgs properties \cite{Aad:2012tfa,Chatrchyan:2012ufa}. These processes also have
importance outside the realm of Higgs physics testing our ability to model isolated photon
radiation in association with strong interactions, see for example Refs.~\cite{Chatrchyan:2013oda,Aad:2013gaa}
for recent experimental studies of the closely related process of photon production in association with hard
jets. From a theoretical point of view di-photon production is
under good control with corrections known up to NNLO in QCD~%
\cite{Catani:2011qz}. NLO QCD corrections to $pp\to\gamma\gamma+j$ have been
available for some time~\cite{DelDuca:2003uz} and have recently been
re-explored~\cite{Gehrmann:2013aga} to investigate the impact of using different photon isolation criteria comparing the smooth
cone isolation~\cite{Frixione:1998jh} with the fixed cone isolation favoured
in experimental studies. The computation of $pp\to\gamma\gamma+2j$ including NLO QCD corrections has been
completed quite recently~\cite{Gehrmann:2013bga,Bern:2013bha}.

In this paper we extend the available range of predictions for $pp\to\gamma\gamma+{}$jets
to include up to three hard jets with NLO accuracy in QCD. The results show a
significant reduction in the uncertainty of the theoretical predictions and
have noticeable corrections to the shapes of the distributions when going from LO to NLO.  As well as
providing new phenomenological studies relevant for the current measurements
at ATLAS and CMS we also present new developments to the \njet{}~C++ code
enabling more efficient computations of high multiplicity processes at NLO.

Modern methods for scattering amplitude computations based on unitarity~\cite{Bern:1994zx},
generalized unitarity~\cite{Britto:2004nc} and integrand
reduction~\cite{Ossola:2006us} have opened up the possibility of performing precision
phenomenological studies with multi-particle final states at colliders. The current
state-of-the-art NLO QCD processes include
$pp\to W/Z+4j$~\cite{Berger:2010zx,Ita:2011wn}, $pp\to 4j$~\cite{Bern:2011ep,Badger:2012pf}, $pp\to W+5j$~\cite{Bern:2013gka}, $pp\to 5j$~\cite{Badger:2013yda},
all of which have been obtained using on-shell
methods. A wealth of $2\to 4$ and general lower multiplicity processes are now
becoming available in an increasing number automated codes \cite{Berger:2008sj,Hirschi:2011pa,Bevilacqua:2011xh,Cullen:2011ac,Cascioli:2011va,Actis:2012qn}.

This article is organized as follows: we begin by outlining our computational
set-up in Section~\ref{sec:setup}, where for completeness we review the well known
decomposition of next-to-leading order differential cross sections and describe
the interface of \njet{} with the Sherpa Monte-Carlo, which we used for the computation of the unresolved real
radiation contributions and phase-space integration. In Section~\ref{sec:results} we provide results for the LHC 
at centre-of-mass energy of $8$~TeV for both $pp\to\gamma\gamma+2j$ and $pp\to\gamma\gamma+3j$.  
We present differential distributions for some important observables, particularly those
used in Higgs productions studies with vector-boson fusion phase space in the case of $pp\to\gamma\gamma+2j$. 
We present a study of the dependence on the renormalization scale of the NLO predictions and investigate the
uncertainty due to the choice of Parton Distribution Functions (PDFs) on total cross sections and distributions.
In Section~\ref{sec:conclusions} we present our conclusions.

\section{Computational set-up \label{sec:setup}}

The computation is performed in the five-flavour scheme with massless b-quark included in the
initial state. The basic partonic sub-processes considered are:
\begin{align*}
  &0 \to \gamma \gamma q\qb gg &
  &0 \to \gamma \gamma q\qb q'\qb'
\end{align*}
for $pp\to \gamma\gamma+2j$ and
\begin{align*}
  &0 \to \gamma \gamma q\qb ggg &
  &0 \to \gamma \gamma q\qb q'\qb'g,
\end{align*}
for $pp\to \gamma\gamma+3j$, from which all relevant channels can be derived using crossing
symmetries. Channels with like-flavour fermions are obtained from the above using the appropriate
(anti-)symmetrization relations.
We do not include loop-induced and formally higher order sub-processes
$0 \to \gamma\gamma + 4g$ and $0 \to \gamma\gamma + 5g$.

We can schematically write down the NLO partonic cross section as a sum of four finite contributions
which can be integrated separately over their respective phase spaces,
\begin{equation}
  d\sigma_n^{NLO} =
    \int_n d\sigma_n^{B} +
    \int_n d\sigma_n^{V} +
    \int_n d\sigma_n^{I} +
    \int_{n+1}d\sigma_{n+1}^{RS},
  \label{eq:NLOdecomp}
\end{equation}
where $d\sigma_n^{B}$ denotes the leading order contribution, $d\sigma_N^{I}$ contains the integrated
dipole subtraction terms, including factorization contributions from initial state singularities, $d\sigma_n^V$
the one-loop virtual corrections and $d\sigma_{n+1}^{RS}$ the infra-red finite contributions from real-radiation
and dipole subtraction terms. This hard partonic cross section is then convoluted with the parton distribution
functions to obtain the cross sections for hadronic collisions.

The computation of the Born and real-emission matrix elements is performed using the colour dressed recursive
Berends-Giele formulation~\cite{Berends:1987me} implemented in the Comix amplitude generator~\cite{Gleisberg:2008fv}.
The subtraction of infra-red singularities is performed using the Catani-Seymour~\cite{Catani:1996jh} dipole method.
The evaluation of these contributions is performed using the Sherpa~\cite{Gleisberg:2007md,Gleisberg:2008ta} package, which is also used
for the determination and organization of the partonic subprocesses and the integration over the phase space.
The one-loop virtual amplitudes are provided using the automated generalized unitarity framework implemented
in the latest version of the \njet{}~library\footnote{The code \njettwo{} is available at \url{https://bitbucket.org/njet/njet/downloads}.}.

\njettwo~is an updated code based on \njet~\cite{Badger:2012pg} and
\ngluon~\cite{Badger:2010nx}. The primitive kinematic objects are constructed
using a numerical generalized unitarity algorithm~\cite{Bern:1994zx,Britto:2004nc,Ossola:2006us,Ellis:2007br,Forde:2007mi,Giele:2008ve,Berger:2008sj,Badger:2008cm}
with Berends-Giele recursion
relations for the tree-level input~\cite{Berends:1987me}. The extended code can compute arbitrary
primitive amplitudes for vector bosons (W, Z and $\gamma$) and massless QCD
partons. Full colour sums are implemented for pure QCD with up to five jets,
vector bosons with up to five jets and di-photon production with up to four
jets. We interface this code to the Sherpa Monte-Carlo via the updated Binoth
Les Houches Accord~\cite{Binoth:2010xt,Alioli:2013nda} to obtain the virtual events. In addition to the standard
internal checks we have managed to check individual phase-space points for the
processes used in this paper against those obtained with {\sc
GoSam}~\cite{Cullen:2011ac} and {\sc MadLoop}~\cite{Hirschi:2011pa}.  For both
processes we have neglected the small effect of top quark loops in the virtual
amplitudes. In the case of $pp\to\gamma\gamma+3j$ we also neglect the
contribution from vector loops where the photons couple directly to a virtual
fermion loop. These corrections have been included in $pp\to\gamma\gamma+2j$
contributing less than 0.5\% of the total cross section, therefore
they are expected to be negligible for $pp\to\gamma\gamma+3j$.

\begin{figure}[h]
  \begin{center}
    \includegraphics[width=0.65\textwidth]{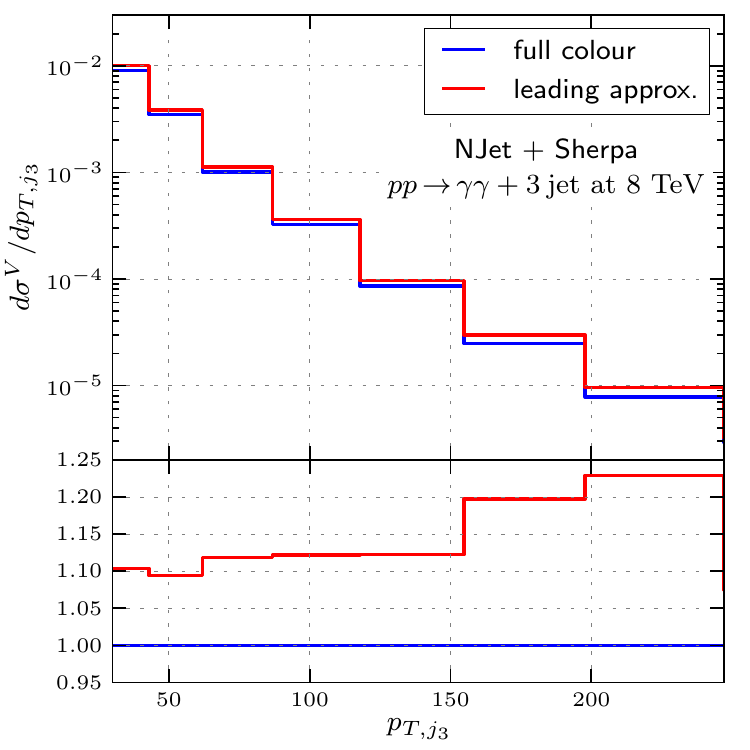}
  \end{center}
  \caption{Full colour and leading approximation (as explained in the text)
  for the virtual corrections to the transverse momentum of the 3rd jet in $pp\to \gamma\gamma+3j$.}
  \label{fig:colour}
\end{figure}

For the current calculation we made use of some new features to optimise the
computation time needed for the virtual corrections. Firstly we use a C++
library Vc~\cite{Kretz_Lindenstruth_2012} to utilize vector capabilities of 
modern CPUs and gain a factor of $\sim2$ in the computation speed.  We also
separate leading and sub-leading contributions in colour such that the simpler,
dominant contributions can be sampled more often than the sub-leading terms.
The definition of our leading terms include all multi-quark processes in the
large $N_c$ limit and processes with two or more gluons in the final state
using the de-symmetrized form of the colour sum that efficiently exploits the
Bose symmetry of the phase space~\cite{Ellis:2009zw,Badger:2012pg}.  The
de-symmetrized sums give full colour information and are faster than leading
colour when we have many final state gluons. In Figure~\ref{fig:colour} we
show virtual corrections to the 3rd leading jet transverse momentum in $pp\to
\gamma\gamma+3j$.  The corrections from the sub-leading part are around $10\%$
on average with a slight rise to around $20\%$ at large $p_T$. In the case of
$pp\to \gamma\gamma+3j$ the virtual cross sections are about $1/3$ of the size
of the total cross section. For the current implementation of $pp\to
\gamma\gamma+3j$, the leading virtual events are generated approximately $7$
times faster than the sub-leading events.

In order to maximise the phenomenological predictions that we can extract from
processes with such complicated final states, we make use of the ROOT Ntuple
format provided by Sherpa~\cite{Binoth:2010ra}. During the course of event
generation, additional weights from the poles in the loop process and from the
subtraction terms are stored along with a full list of kinematic variables and
couplings. This allows the events to be re-weighted to different scale and PDF
choices during analysis.  Even within this approach a full
study of PDF uncertainties and scale variations can be computationally
intensive. During this work we have developed an interface to the
APPLgrid library~\cite{Carli:2010rw} to allow for specific observables to be
efficiently re-weighted changing the scales and the specific PDF set used.

\section{Numerical results \label{sec:results}}

All the results presented in this section are for $pp$ collisions with a centre-of-mass energy of
8~TeV. We consider the following kinematic cuts on the external momenta, which are inspired by
typical experimental cuts used in the analyses at LHC
\begin{align*}
  &p_{T,j} > 30 \,\GeV && |\eta_j|\leq 4.7 && \\
  &p_{T,\gamma_1} > 40 \,\GeV && p_{T,\gamma_2} > 25 \,\GeV && |\eta_\gamma|\leq 2.5 \\
  &R_{\gamma,j} = 0.5 && R_{\gamma,\gamma} = 0.45
\end{align*}
where the photon transverse momenta have been ordered by size. The jets are defined using the
anti-$k_T$ algorithm \cite{Cacciari:2008gp} with cone size $R=0.5$ as implemented in {\sc FastJet}~\cite{Cacciari:2011ma}.
Photons are selected using the Frixione smooth cone isolation criterion~\cite{Frixione:1998jh}. A photon
is considered isolated if the total hadronic energy inside all cones of radius $r_\gamma < R$
\begin{equation}
  E_{\text{hadronic}}(r_\gamma) \leq \eps \,p_{T,\gamma} \left( \frac{1-\cos{r_\gamma}}{1-\cos{R}} \right)^n
  \label{eq:frixionecone}
\end{equation}
with $\eps=0.05$, $R=0.4$ and $n=1$. We use the NLO CT10 PDF set~\cite{Lai:2010vv} for our central predictions
with the strong coupling running from $\alpha_s(M_Z) = 0.118$, and the electromagnetic coupling fixed
at $\alpha=1/137.036$. In particular we use the same (NLO) PDF set and definition of the strong coupling constant
both for LO and NLO predictions. Using a NLO PDF set for the LO computation includes higher order terms that go
beyond a fixed order prediction, nevertheless such a set-up allows us
to separate NLO effects coming from the running of the strong coupling and from
PDFs and to highlight the impact of corrections coming from the NLO matrix
elements.

We choose a dynamical value for the factorization and renormalization scales which are kept equal,
$\mu_R = \mu_F$, when performing scale variations. We have investigated the dependence on a number of different 
functional forms which we will denote as:
\begin{align}
  \widehat{H}_T &= p_{T,\gamma_1} + p_{T,\gamma_{2}} + \sum_{i\in\text{partons}} p_{T,i} \\
  \widehat{H}_T' &= m_{\gamma\gamma}  + \sum_{i\in\text{partons}} p_{T,i} \\
  \widehat{\Sigma}^2 &= m_{\gamma\gamma}^2  + \sum_{i\in\text{partons}} p_{T,i}^2 \\
  H_T' &= m_{\gamma\gamma}  + \sum_{i\in\text{jets}} p_{T,i} \\
  \Sigma^2 &= m_{\gamma\gamma}^2  + \sum_{i\in\text{jets}} p_{T,i}^2
  \label{eq:scaledefs}
\end{align}
where $m_{\gamma\gamma} = \sqrt{p_{T,\gamma_1}^2+p_{T,\gamma_2}^2}$. The quantities $H_T'$ and
$\Sigma^2$ are constructed after the clustering of final state partons into jets. Notice that partonic and jet-level
 scales will only differ at NLO, where the additional unresolved radiation enters in the clustering algorithm.

\subsection{Results for $pp\to \gamma\gamma+2j$}

\begin{figure}[t]
  \begin{center}
    \includegraphics[width=0.8\textwidth]{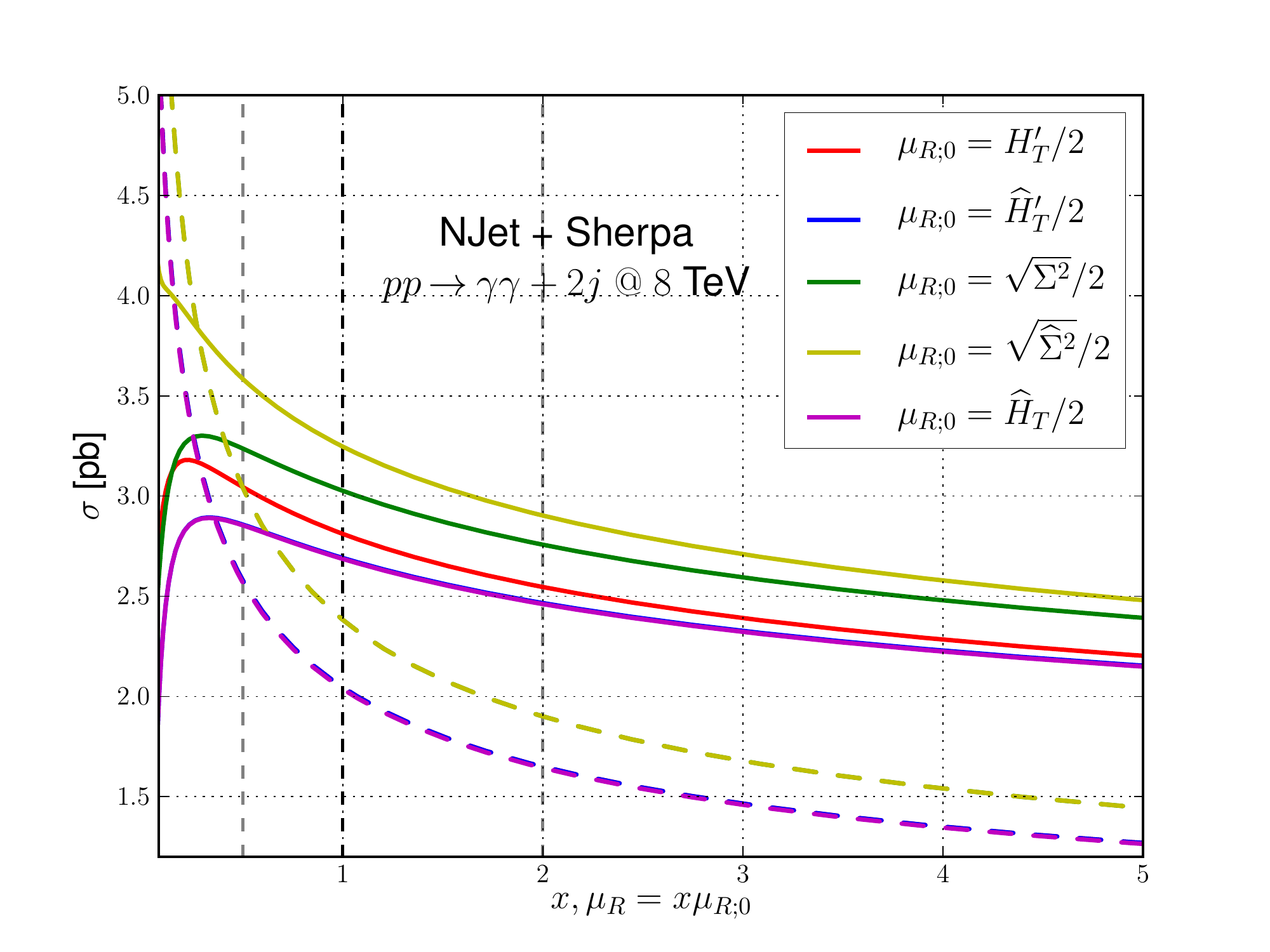}
  \end{center}
  \caption{Scale variations on the total cross section for $pp\to \gamma\gamma+2j$ for a variety
  of dynamical scales. Dashed lines are LO accuracy whereas solid lines are NLO accuracy. The vertical
  line at $x=1$ corresponds to the central scale whereas the lines at $x=0.5$ and $x=2$ show the bounds
  of the scale variation region.}
  \label{fig:AA2j_scalevar}
\end{figure}

We first consider the production of a photon pair in association with two jets. In this case
we compare our predictions with the recent results of reference~\cite{Gehrmann:2013bga}
and present additional studies of PDF variations and dynamical scale choices.
For the latter we rely on the possibility to substantially reduce the computational cost
by the use of our APPLgrid set-up.

In Figure~\ref{fig:AA2j_scalevar} we show the dependence of the total inclusive cross section
upon variation of the renormalization and factorization scales with $\mu_R=\mu_F$. We consider the five different
dynamical scales defined in Eq.~\eqref{eq:scaledefs}. Though the scale choices are closely related to each other
we find significant deviations for the value of the total cross section. Nevertheless the NLO
predictions show a significant reduction in the dependence on the scale variation compared to the
LO ones. Taking the envelope of all the scales considered we see that the LO predictions vary in the interval
$1.64-3.04$ whereas NLO ones lie within $2.46-3.58$ when the scales are varied over the range
$x\in[0.5,2]$ around the central choice. This represents a reduction in the scale variation uncertainty
from $\sim 30\%$ at LO to $\sim 20\%$ at NLO.

We notice that choosing $\mu_R=\sqrt{\widehat{\Sigma}^2}$ as a scale leads to significantly larger
predictions for the total cross section than with the other scales considered here. Moreover the scale variation
profile closely resembles the one of a leading order prediction. We therefore consider this choice of scale
disfavoured with respect to the others, which provide results that are in better agreement with each other.
In general, we find that the $H_T$-based scales lead to a broader profile of scale variations and on average
favour harder events than the $\Sigma$-based scales. On the other hand we find the shapes of the distributions 
to be quite stable with respect to the scale choice.

The results for the total cross section and distributions using $\mu_R=\sqrt{\Sigma^2}/2$ are
in good agreement with those obtained previously by Gehrmann, Greiner and Heinrich~\cite{Gehrmann:2013bga}.
In the following we opt for the scale $\widehat{H}_T'$ for computing our predictions for the total cross section and
the differential distributions presented here. This scale has been widely used in studies of $W+{}$jets (see for example~\cite{Bern:2013gka}).

The values for the total cross sections at both LO and NLO computed our default choices for scale, PDF set and physical
parameters are found to be:
\begin{align}
  \sigma_{\gamma\gamma+2j}^{LO}(\widehat{H}_T'/2) &= 2.046(0.002)^{+0.534}_{-0.396}\,{\rm pb} &
  \sigma_{\gamma\gamma+2j}^{NLO}(\widehat{H}_T'/2) &= 2.691(0.007)^{+0.167}_{-0.225}\,{\rm pb}
  \label{eq:AA2jinclXS}
\end{align}
where the sub-scripts(super-scripts) show the maximum deviation from the central value in the negative(positive) direction
over the range $x\in[0.5,2]$ for $\mu_R = x \widehat{H}_T'/2$. Monte-Carlo integration errors are shown in brackets.

\begin{figure}[h]
  \begin{center}
    \includegraphics[width=0.45\textwidth]{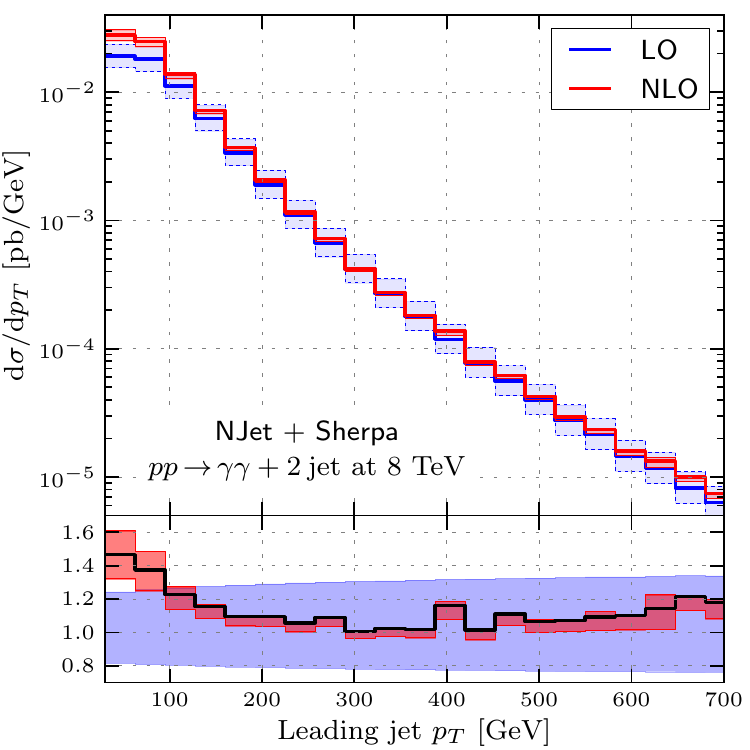}
    \includegraphics[width=0.45\textwidth]{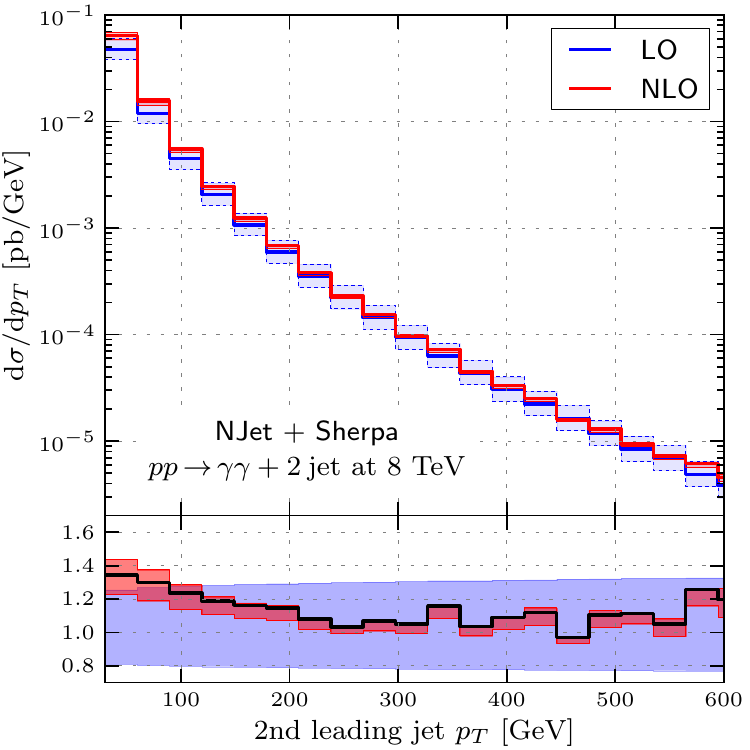}
  \end{center}
  \caption{Differential distributions for jet transverse momenta. The lower plot shows the ratio
  of NLO to LO including the scale variation bands estimated over the range of $x\in[0.5,2]$.}
  \label{fig:aa2j_jet_pt}
\end{figure}

\begin{figure}[h]
  \begin{center}
    \includegraphics[width=0.45\textwidth]{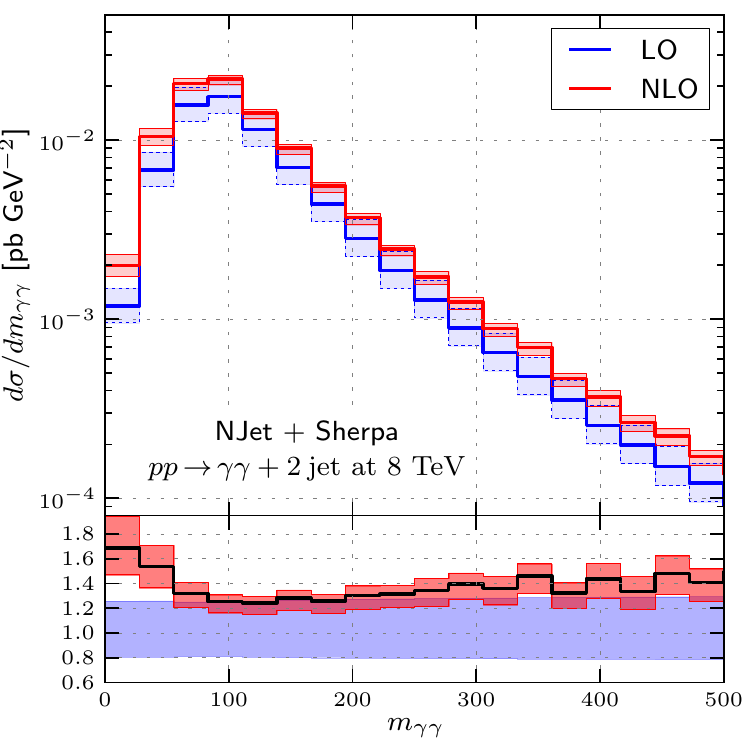}
    \includegraphics[width=0.45\textwidth]{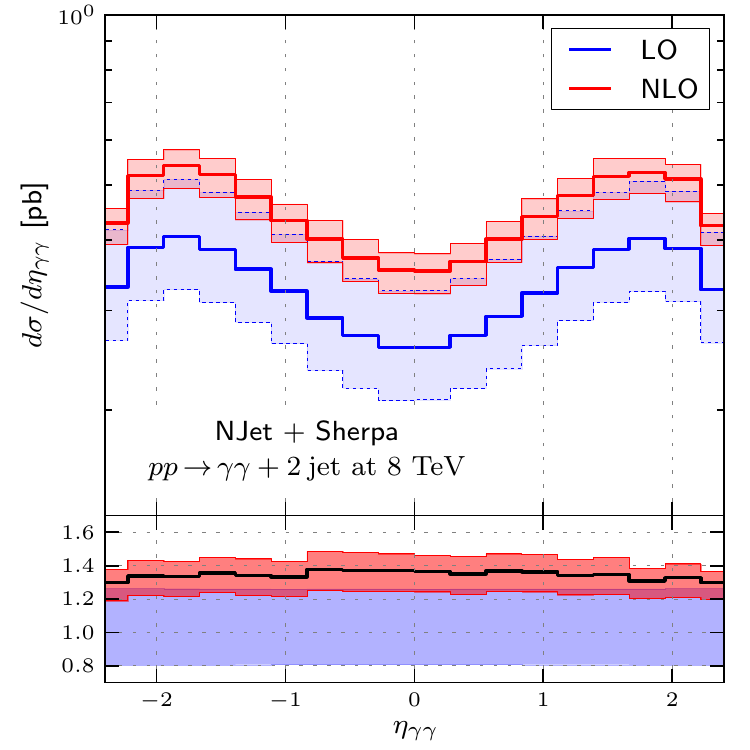}
  \end{center}
  \caption{Differential distributions for di-photon invariant mass and rapidity $pp\to\gamma\gamma+2j$.}
  \label{fig:aa2j_photonpair}
\end{figure}

Figure \ref{fig:aa2j_jet_pt} shows the differential distributions for the ordered jet transverse momenta.
The results for the scale $\widehat{H}_T'/2$ are consistent with those obtained at the scale $\sqrt{\Sigma^2}$
with a significant reduction in scale variation from around $20\%$ at LO to $\sim 10\%$ at NLO in both cases.
The $K$-factor is fairly constant at around $1.1$ for $p_T$ higher than 200 GeV rising to $1.4$ as we approach the $p_T$ cut.
This larger $K$-factor in the low $p_T$ region could be the indication of the presence of large logarithms beyond fixed order NLO.
The di-photon invariant mass $m_{\gamma\gamma}$ and the di-photon rapidity distributions $\eta_{\gamma\gamma}$ are shown in
Figure~\ref{fig:aa2j_photonpair}. They receive slightly larger NLO corrections with respect to the jet transverse momenta with
the $K$-factor for the former ranging from $1.2$ in the large invariant mass region to $1.7$ for lower invariant masses, while
for the latter NLO correction induce a roughly flat $K$-factor of~$1.3$.

\begin{figure}[h]
  \begin{center}
    \includegraphics[width=0.45\textwidth]{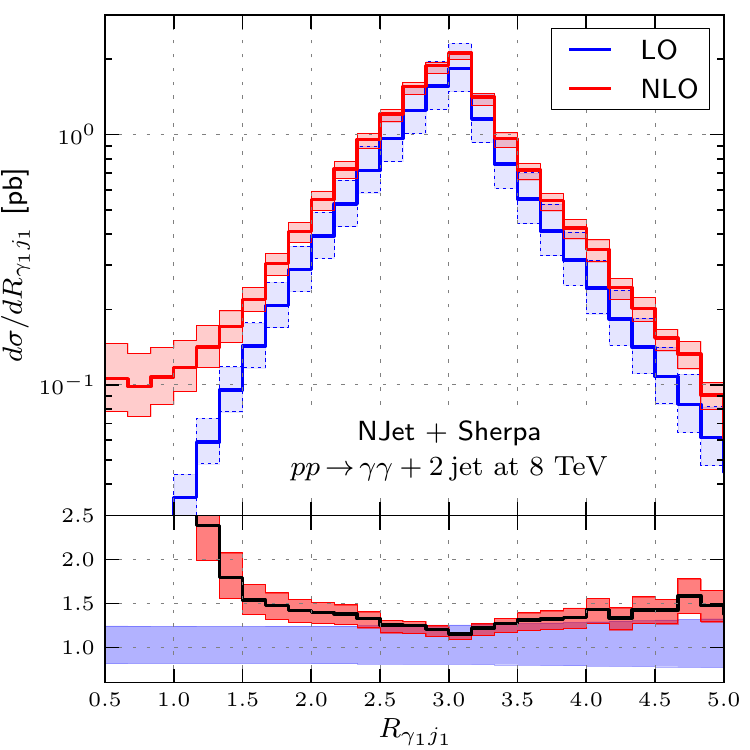}
    \includegraphics[width=0.45\textwidth]{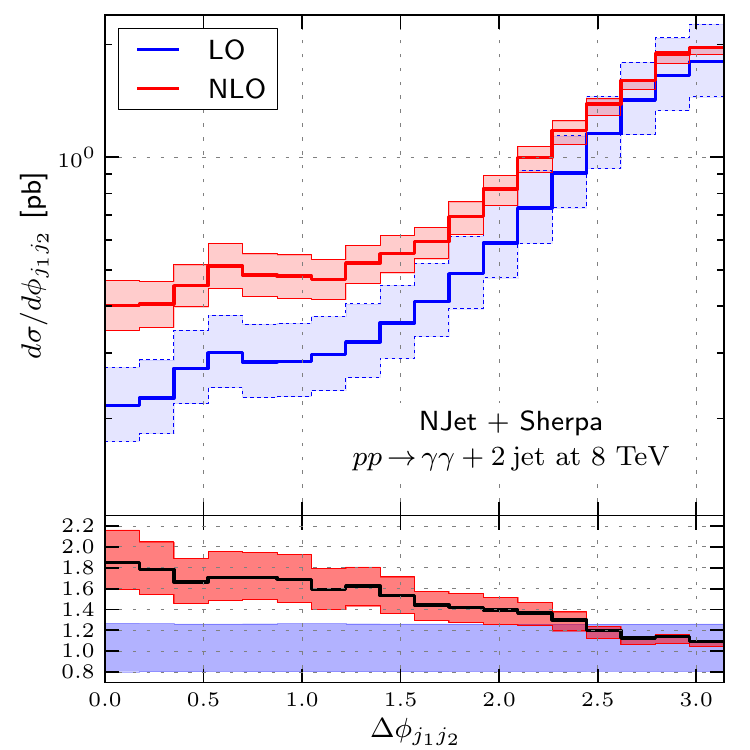} \\
    \includegraphics[width=0.45\textwidth]{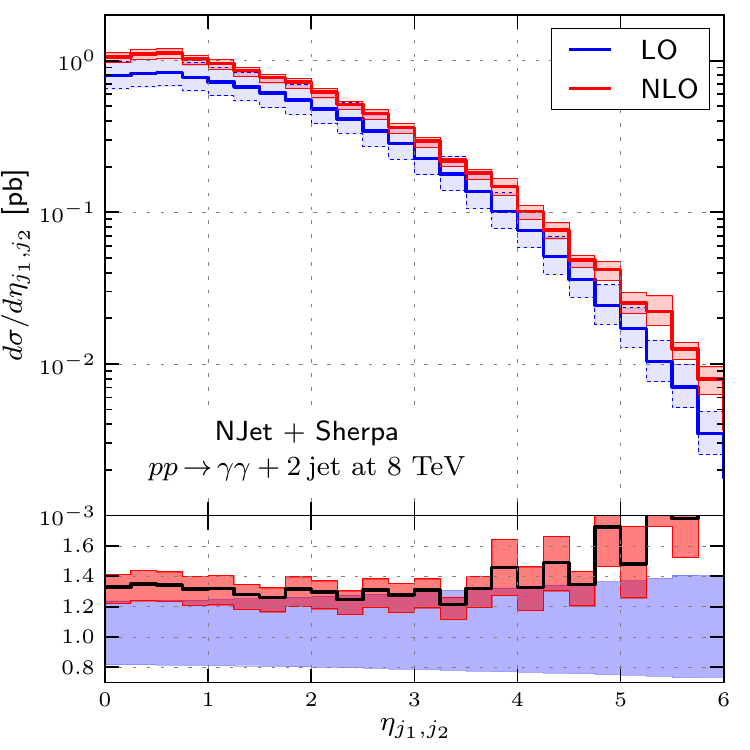}
    \includegraphics[width=0.45\textwidth]{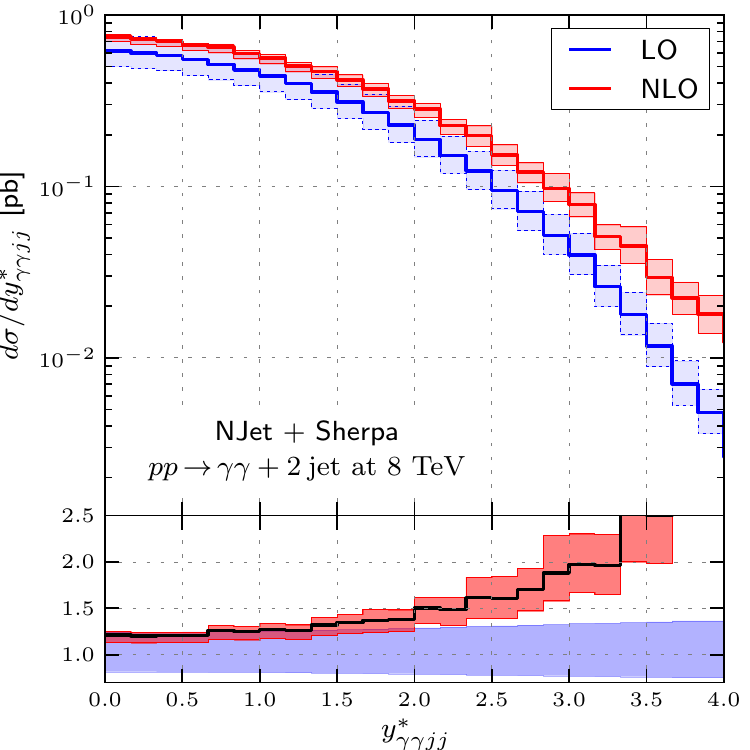}
  \end{center}
  \caption{Differential distributions for the angular observables $R_{\gamma_1 j_1}$ (see Eq.~\eqref{eq:R11}),
  $\Delta\phi_{j_1j_2}$, $\eta_{j_1 j_2}$ and $y_{\gamma\gamma jj}^*$ (see Eq.~\eqref{eq:ystar}) in $pp\to\gamma\gamma+2j$.}
  \label{fig:aa2j_vbf}
\end{figure}

Figure \ref{fig:aa2j_vbf} shows four distributions of angular quantities that are usually used in analyses of 
Higgs production in vector boson fusion (VBF), where additional cuts are imposed in order to reduce QCD 
backgrounds in $pp\to H (\to\gamma\gamma)+2j$ studies.

Owing to the increased phase-space available to the real radiation at NLO we see large deviations
from the shapes of the leading order distributions. These features have been pointed out for the jet-pair
azimuthal angle $\Delta\phi_{j_1j_2}$ and for the separation of the leading-photon/leading jet,
\begin{align}
  R_{\gamma_1 j_1} = \sqrt{\Delta y_{\gamma_1 j_1}^2 +  \Delta\phi_{\gamma_1 j_1}^2},
  \label{eq:R11}
\end{align}
in Ref.~\cite{Gehrmann:2013bga}. We also see increasing deviations for large rapidities of the jet pair $\eta_{j_1 j_2}$
and even more so in the relative rapidity of the diphoton-dijet system,
\begin{align}
  y_{\gamma\gamma jj}^* = y_{\gamma\gamma} - (y_{j_1}+y_{j_2})/2.
  \label{eq:ystar}
\end{align}

Using the APPLgrid set-up described in the previous section, we have also performed a study of PDF uncertainties on
$pp\to\gamma\gamma+2j$, concentrating on the total cross section and the invariant mass distribution of the photon
pair. In Table \ref{tab:aa2jXSpdf} we show the central value and PDF uncertainties for the total cross section evaluated
at the central scale ($\mu_R=\widehat{H}_T'/2$) using four different NLO PDF sets: CT10~\cite{Lai:2010vv},
MSTW2008~\cite{Martin:2009iq}, ABM11~\cite{Alekhin:2012ig} and NNPDF2.3~\cite{Ball:2012cx}. All PDF sets are compared
using the same value of $\alpha_s(M_Z) = 0.118$ in order to disentangle PDF and strong coupling constant uncertainties.
PDF uncertainties are considerably smaller than the theoretical uncertainty estimated from scale variations and range from
1\% for MSTW and NNPDF to 3.5\% for ABM. We note, however that the ABM11 uncertainty includes errors associated with
$\alpha_s$ variations. In general, we find that central predictions from different PDF sets differ by amounts which are larger
than the nominal PDF uncertainty of each set.

\begin{table}
  \centering
  \renewcommand\arraystretch{1.3}
  \begin{tabular}{lccc}
      \hline
    PDF set & $\sigma_{\gamma\gamma+2j}^{NLO}(\widehat{H}_T'/2)$
      & $\delta\sigma_{\gamma\gamma+2j}^{NLO,PDF+}(\widehat{H}_T'/2)$
      & $\delta\sigma_{\gamma\gamma+2j}^{NLO,PDF-}(\widehat{H}_T'/2)$ \\
      \hline
      CT10nlo   & $2.69102$ & $+0.0357456$ & $-0.042148$ \\
      NNPDF2.3  & $2.77285$ & $+0.0167702$ & $-0.016770$ \\
      MSTW2008  & $2.71578$ & $+0.0184072$ & $-0.016373$ \\
      ABM11     & $2.73791$ & $+0.0659662$ & $-0.065966$ \\
      \hline
  \end{tabular}
  \caption{The total cross section for $pp\to \gamma\gamma+2j$ computed with different PDF sets at $\alpha_s(M_Z)=0.118$.
  	The PDF uncertainties are computed from the relevant PDF error sets.
  }
  \label{tab:aa2jXSpdf}
\end{table}

\begin{figure}[h]
  \begin{center}
    \includegraphics{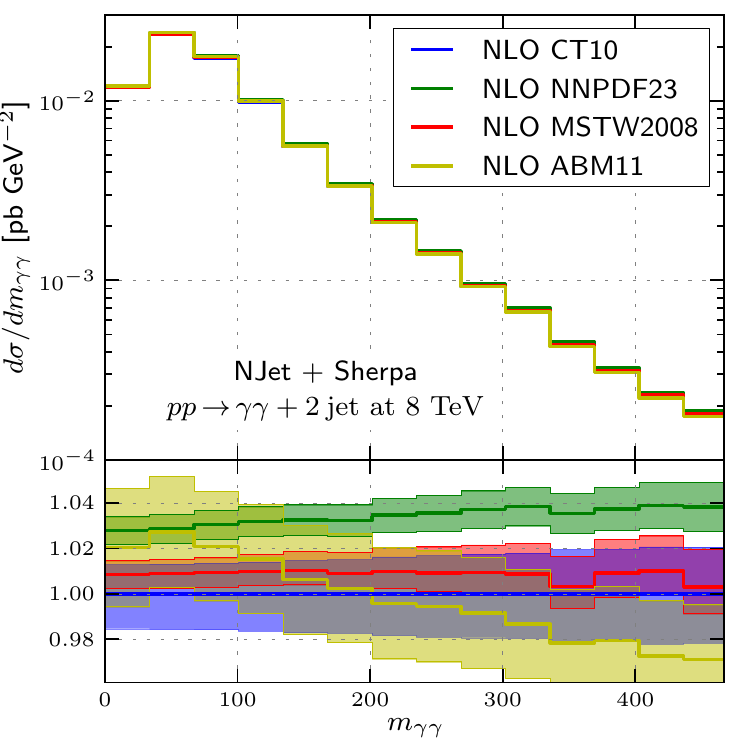}
  \end{center}
  \caption{The $m_{\gamma\gamma}$ distributions for $pp\to \gamma\gamma+2j$ for the four PDF sets described in the text
  at $\alpha_s{M_Z} = 0.118$. The lower plot shows the ratio of each set to CT10 with the shaded region representing
  the PDF uncertainty.}
  \label{fig:AA2jmaadistPDF}
\end{figure}

In Figure \ref{fig:AA2jmaadistPDF} we show the distributions for the invariant mass of the photon pair ($m_{\gamma\gamma}$)
computed at the scale $\mu_R=\widehat{H}_T'/2$ and the four PDF sets considered before at $\alpha_s(M_Z) = 0.118$.
The upper (log scale) plot shows the absolute predictions for all sets, which are indeed extremely
close to each other. In the lower plot, where we show the ratio of each set to the central value
computed using CT10, we do however notice discrepancies in the predictions of the order of a few percent.
These differences are especially pronounced at high $m_{\gamma\gamma}$ where the spread between them reaches $\sim 6\%$, 
a value larger than PDF uncertainties coming from individual sets (represented by the shaded regions). We also note that, while 
the differences between CT10, MSTW08 and NNPDF2.3 are mostly due to an overall normalization factor, the ABM11 predictions 
stand out from the others also in the shape being higher for low values of $m_{\gamma\gamma}$ and dropping below the other 
predictions for higher values of the invariant mass.

\subsection{Results for $pp\to \gamma\gamma+3j$}

\begin{figure}[h]
  \begin{center}
    \includegraphics[width=0.8\textwidth]{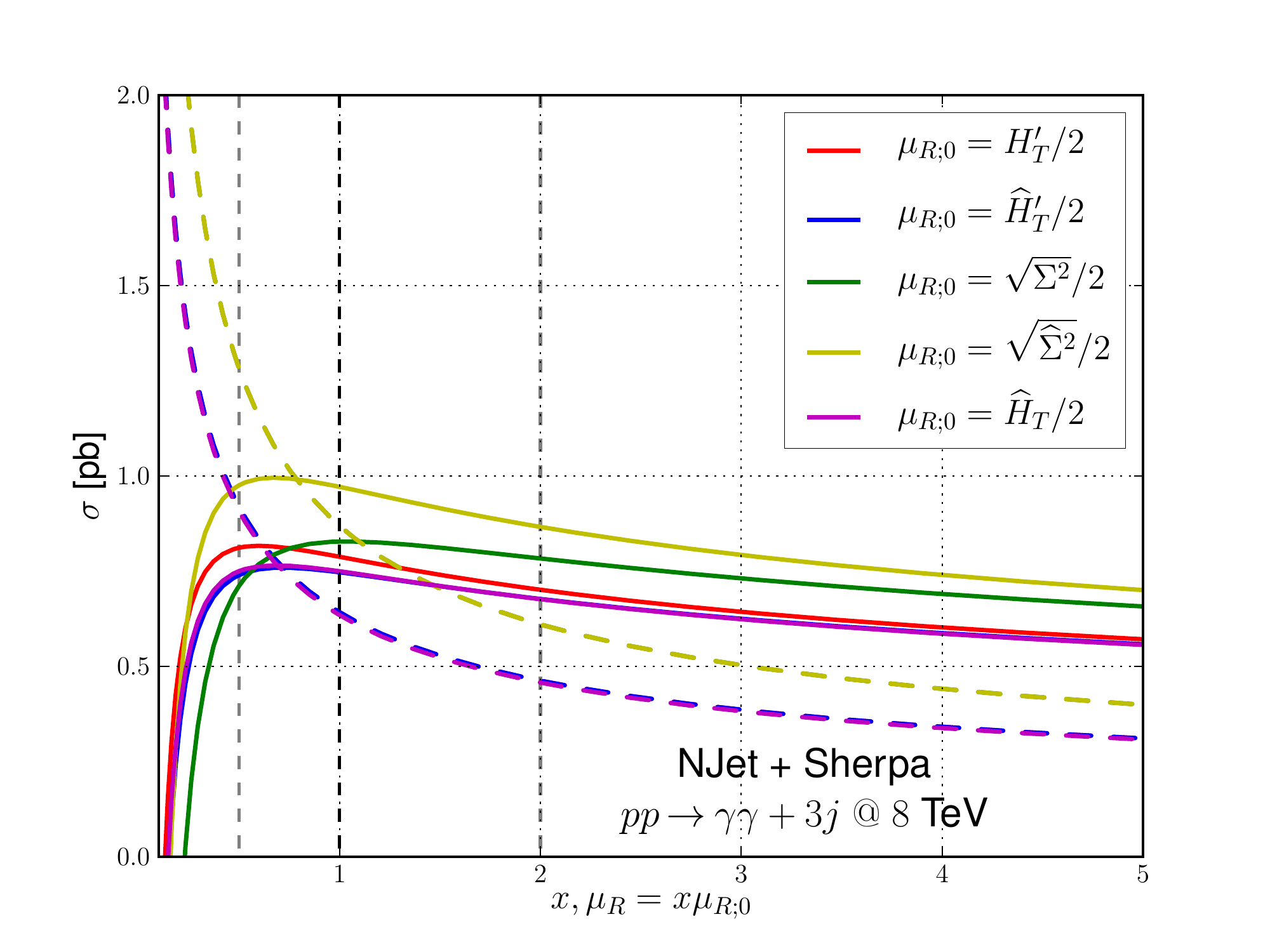}
  \end{center}
  \caption{The result of scale variations on the total inclusive cross section for
  $pp\to\gamma\gamma+3j$. LO~curves are represented with dashed lines while NLO
  curves are represented with solid lines.}
  \label{fig:AA3j_scalevar}
\end{figure}

We now consider the production of a photon pair in association with three jets.
As in the previous section we studied the dependence
of the total cross section upon variation of renormalization and factorization
scales with the choices of dynamical scales defined in Eq.~\eqref{eq:scaledefs}.
The results in Figure~\ref{fig:AA3j_scalevar} show reasonable
differences between quantities based on jets versus quantities based on
partons.  Overall we find a significant improvement in the uncertainty
estimated from scale variations when going from LO to NLO. The envelope of
predictions from all scale choices varied over the range $x\in[0.5,2]$ is
around $0.67-0.99$~pb at NLO compared to $0.46-1.28$~pb at LO. This represents a
decrease in variation from $\sim 50\%$ at LO to $\sim 20\%$ at NLO. As in the two
jet case the scales based on $\Sigma^2$ give generally larger predictions than
those based on $H_T$. Other than the overall normalization, we find that all
scales give very similar predictions for shapes of the distributions.

Comparing Figure~\ref{fig:AA3j_scalevar} with Figure~\ref{fig:AA2j_scalevar} we see that
the peak in the NLO curve for $\Sigma^2$ has moved further to the right than the
$H_T$ scales which may suggest that a range of $x\in[1,4]$ would be more appropriate
here. Since we would like to make predictions for jet ratios we need to have
as consistent description of $\gamma\gamma+3j$ and $\gamma\gamma+2j$ as
possible and therefore we prefer the $H_T$ scales. In the following we choose to adopt
the central scale of $\widehat{H}_T'/2$ for the total rates and distributions, though theoretical 
uncertainties are likely underestimated by the simple scale variations following the discussion above. 
For the total cross sections at LO and NLO we find,
\begin{align}
  \sigma_{\gamma\gamma+3j}^{LO}(\widehat{H}_T'/2) &= 0.643(0.003)^{+0.278}_{-0.180}\,{\rm pb} &
  \sigma_{\gamma\gamma+3j}^{NLO}(\widehat{H}_T'/2) &= 0.785(0.010)^{+0.027}_{-0.085}\,{\rm pb}
  \label{eq:AA3jinclXS}
\end{align}
where the sub-scripts(super-scripts) show the maximum deviation from the
central value in the negative(positive) direction over the range $x\in[0.5,2]$
for $\mu_R = x \widehat{H}_T'/2$ and Monte-Carlo integration errors are shown in
brackets.

\begin{figure}[h]
  \begin{center}
    \includegraphics[width=0.45\textwidth]{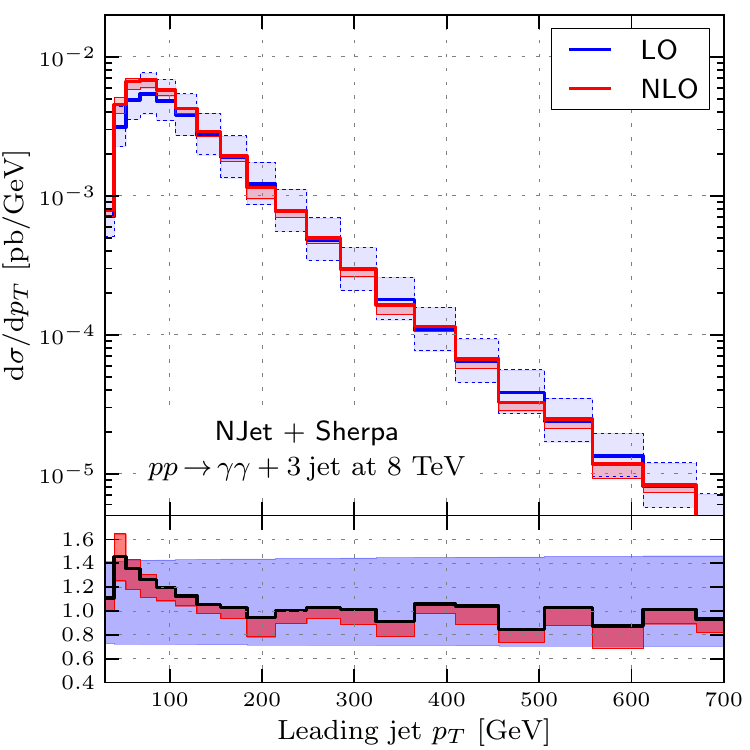}
    \includegraphics[width=0.45\textwidth]{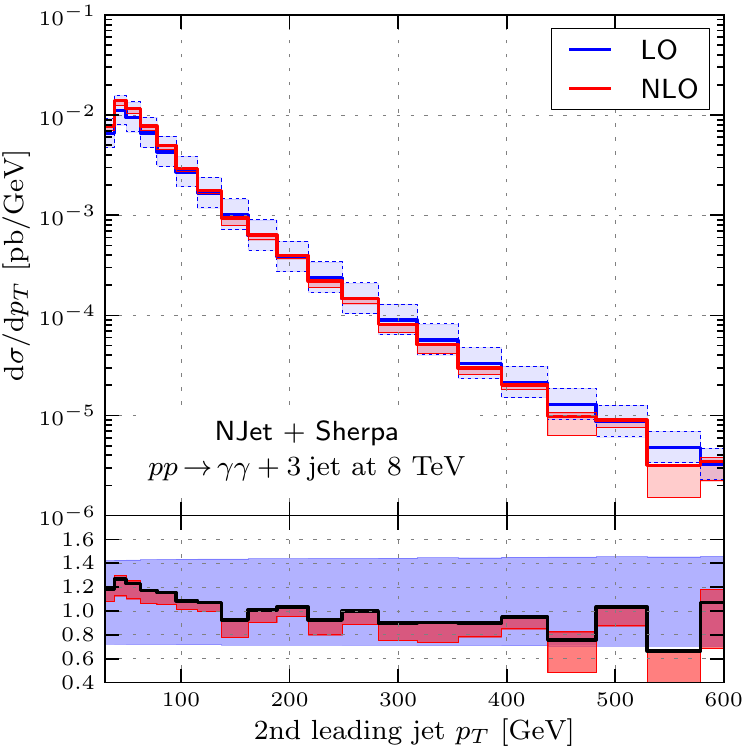}
  \end{center}
  \caption{Differential cross section as a function of $p_T$ of the 1st and 2nd leading jets.}
  \label{fig:AA3j_jpt}
\end{figure}

\begin{figure}[h]
  \begin{center}
\subfloat[]{\label{fig:AA3j_jpt3}%
    \includegraphics[width=0.45\textwidth]{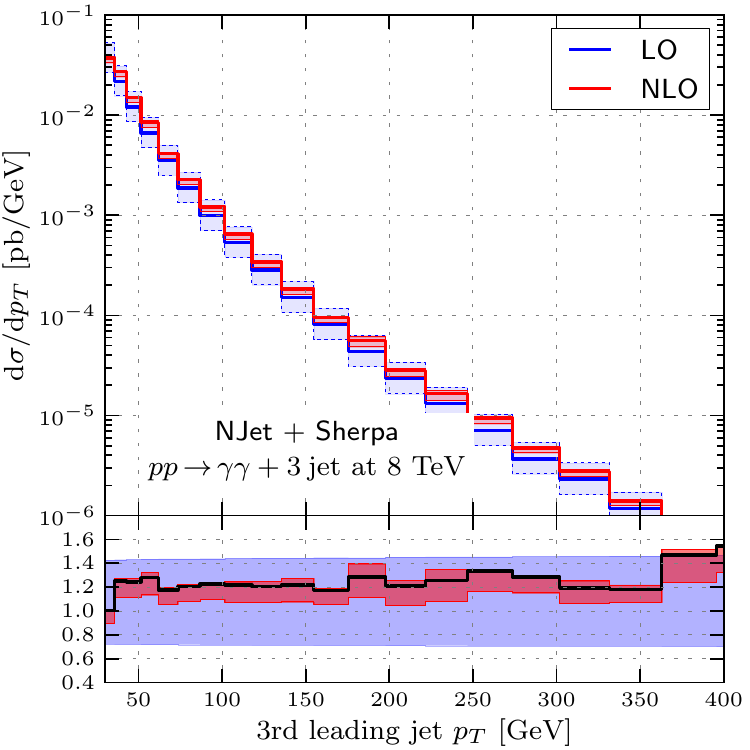}}
\subfloat[]{\label{fig:photon_mass}%
    \includegraphics[width=0.45\textwidth]{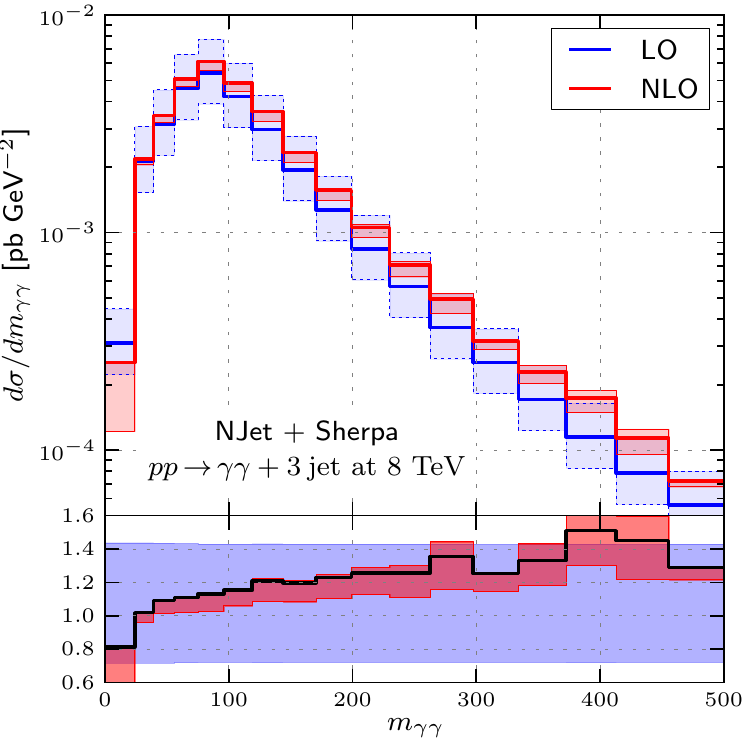}}
  \end{center}
  \caption{Differential cross section as a function of $p_T$ of the 3rd leading jet and
           di-photon invariant mass in $pp\to \gamma\gamma+3j$.}
\end{figure}

The distributions for the jets transverse momenta are shown in Figure~\ref{fig:AA3j_jpt}. The
$K$-factor is quite flat with a value between $1.0$ and $1.2$ except in the
low $p_T$ where it rises to around $1.4$ for the leading jet $p_T$. Again, this
may suggest the presence of large logarithms in the missing higher order
contributions in this region. The distribution of the 3rd jet shown in Figure~\ref{fig:AA3j_jpt3} has a noticeably
flatter $K$-factor than the ones for the two leading jets.

\begin{figure}[h]
  \begin{center}
    \includegraphics[width=0.45\textwidth]{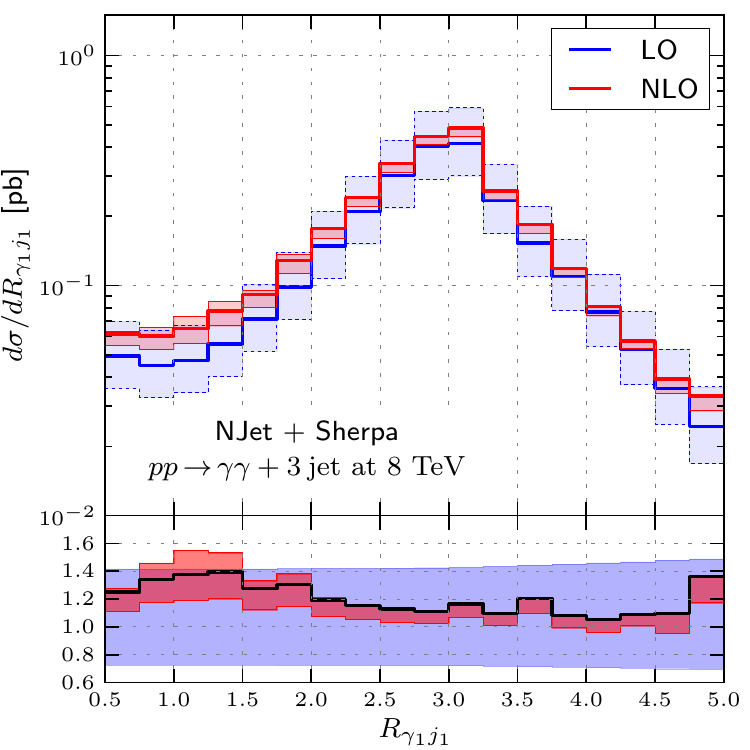}
    \includegraphics[width=0.45\textwidth]{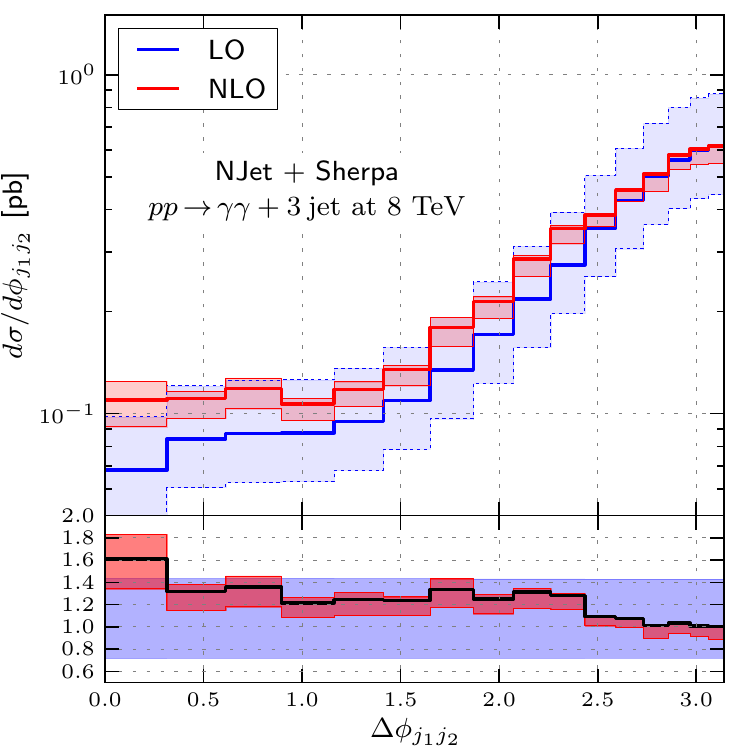}
  \end{center}
  \caption{Photon/jet separation and di-jet azimuthal angle distributions in $pp\to \gamma\gamma+3j$.}
  \label{fig:extra_dist}
\end{figure}

The di-photon invariant mass distribution in Figure~\ref{fig:photon_mass} receives significant corrections to
the shape at NLO with the $K$-factors increasing from around $1.0$ at low $m_{\gamma\gamma}$ to $1.4$ at
large values of the photon pair invariant mass. In Figure~\ref{fig:extra_dist} we show the leading jet/leading
photon separation $R_{\gamma_1 j_1}$ and the azimuthal separation of the two leading jets $\Delta\phi_{j_1 j_2}$.
Both quantities receive large NLO corrections for small values of the observable, though notably not as large as
the corresponding distributions in $pp\to\gamma\gamma+2j$ where there is a more substantial increase in the  
available phase-space at NLO.

\begin{figure}[h]
  \begin{center}
    \includegraphics{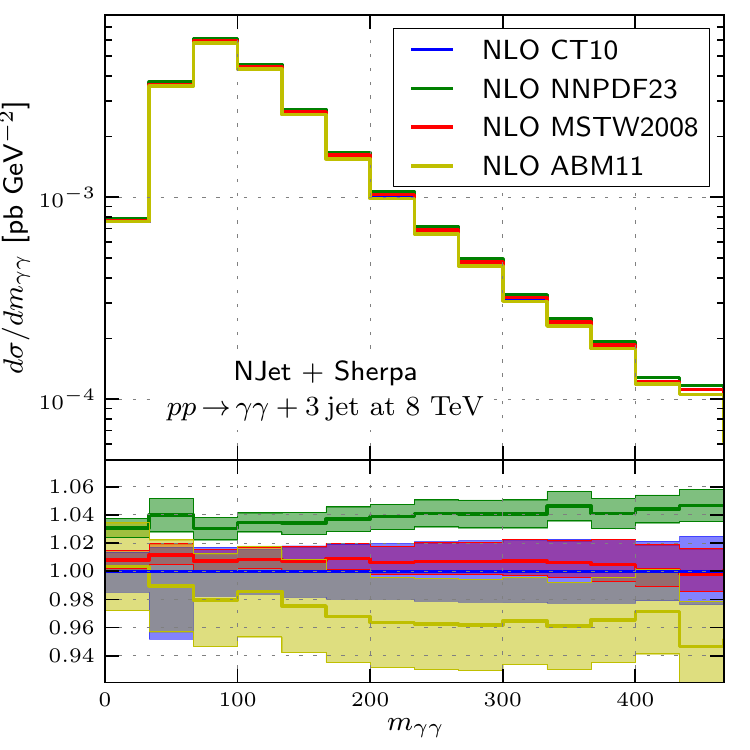}
  \end{center}
  \caption{The $m_{\gamma\gamma}$ distributions for $pp\to \gamma\gamma+3j$ for the four PDF sets described in the text
  at $\alpha_s{M_Z} = 0.118$. The lower plot contains the ratio of each set to CT10 with the shaded region representing
  the PDF uncertainties.}
  \label{fig:AA3jmaadistPDF}
\end{figure}

The PDF analysis was performed with the same set-up as for $pp\to \gamma\gamma+2j$ with the
four PDF sets compared at the same value of $\alpha_s(M_Z)=0.118$. Table~\ref{tab:aa3jXSpdf}
shows the results for the total cross section at the central scale $\mu_R = \widehat{H}_T'/2$.
Again we see that PDF uncertainties for all sets are noticeably smaller than theoretical uncertainties
estimated from scale variations and that central predictions from different sets differ by more than the
nominal uncertainty obtained with each set.
Figure~\ref{fig:AA3jmaadistPDF} shows the comparison of the $m_{\gamma\gamma}$ distribution
for the different PDF~sets. The deviation between ABM11 and NNPDF, which again give the lower and higher limit
of the predictions from different sets, is somewhat larger than that seen in $pp\to \gamma\gamma+2j$,
though the results are consistent within the theoretical uncertainties determined by scale variations.

\begin{table}
  \centering
  \renewcommand\arraystretch{1.3}
  \begin{tabular}{lccc}
      \hline
    PDF set & $\sigma_{\gamma\gamma+3j}^{NLO}(\widehat{H}_T'/2)$
      & $\delta\sigma_{\gamma\gamma+3j}^{NLO,PDF+}(\widehat{H}_T'/2)$
      & $\delta\sigma_{\gamma\gamma+3j}^{NLO,PDF-}(\widehat{H}_T'/2)$ \\
      \hline
      CT10nlo   & $0.746696$ & $+0.0123788$ & $-0.0133826$ \\
      NNPDF2.3  & $0.773112$ & $+0.0056425$ & $-0.0056425$ \\
      MSTW2008  & $0.752756$ & $+0.0068782$ & $-0.0050721$ \\
      ABM11     & $0.731019$ & $+0.0241568$ & $-0.0241568$ \\
      \hline
  \end{tabular}
  \caption{The total cross section for $pp\to \gamma\gamma+3j$ computed with different
  PDF sets at $\alpha_s(M_Z)=0.118$. The PDF uncertainties are computed from the relevant PDF error sets.
  }
  \label{tab:aa3jXSpdf}
\end{table}

\subsection{The three-to-two jet ratio}

In the context of multi-jet production studies it is interesting to look at the ratio of $pp\to \gamma\gamma+3j$
over $pp\to \gamma\gamma+2j$ which we will denote as $R_{3/2}$.
Due to the cancellation of many uncertainties both theoretical and experimental, ratios such as this one are prime
observables for the determination of physical parameters (like $\alpha_s$).

\begin{figure}[h]
  \begin{center}
    \includegraphics{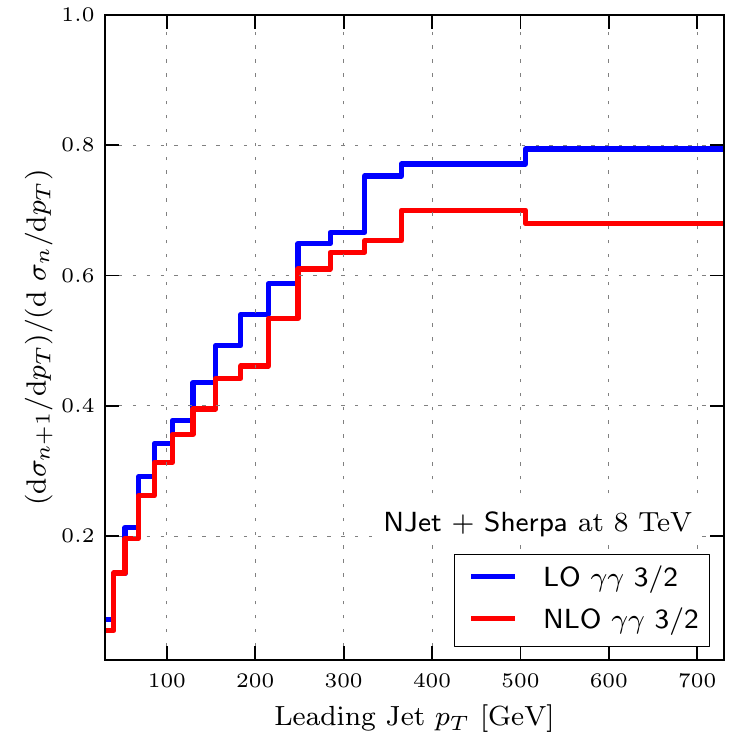}
  \end{center}
  \caption{The ratio of of $pp\to \gamma\gamma+3j$ over $pp\to \gamma\gamma+2j$ as a function of leading jet~$p_T$.}
  \label{fig:jetratio_pT_1}
\end{figure}

In the case of the production of di-photon in association with jets, we find $R_{3/2}$ to be:
\begin{align}
  R_{3/2}^{LO}(\mu_R=\widehat{H}_T'/2) &= 0.314(0.002) &
  R_{3/2}^{NLO}(\mu_R=\widehat{H}_T'/2) &= 0.276(0.004)
  \label{eq:3to2ratio}
\end{align}
where the numbers in brackets refer to Monte-Carlo errors. We have checked that all scale choices
are in much better agreement for $R_{3/2}$ with the range of predictions lying within $\sim 8\%$
of this value. In Figure \ref{fig:jetratio_pT_1} we show the differential ratio with the $p_T$ of the
leading jet. The NLO corrections become more important for $p_T>100$ GeV and reaching
about $15\%$ at high $p_T$.

\section{Conclusions \label{sec:conclusions}}

In this paper we have presented a study of di-photon production in association with jets.
The first calculation of the full NLO QCD corrections to the process $pp\to\gamma\gamma+3j$ is presented and discussed,
together with a variety of results for  $pp\to\gamma\gamma+2j$. We find that the inclusion of NLO QCD corrections leads
to a significant reduction of theoretical uncertainties both on total cross sections and distributions.
We have studied distributions for a number of observables, of particular interest are those relevant for Higgs production in VBF
analyses which are used when modelling $\gamma\gamma+{}$jets as a background to $pp\to H+\text{jets}\to \gamma\gamma+{}$jets.

The present study is based on the use of the Frixione smooth cone isolation criterion to define the final state photons, which
provides a theoretically clean way of suppressing the fragmentation component in direct photon production. It would be
also interesting to consider alternative isolation criteria, like the fixed cone isolation, which require the inclusion of fragmentation
functions, and were shown to have a significant effect in lower multiplicity processes \cite{Gehrmann:2013aga}.
Moreover the majority of experimental analyses involving direct photon production (both in association with jets or not) rely
on the cone isolation for photon identification.
Nonetheless, we hope that the results presented here will be of use in future experimental analyses and look forward
to direct comparisons with the LHC data.

\acknowledgments{%
We would like to thank Peter Uwer, Benedikt Biedermann, Thomas Gehrmann, Gudrun Heinrich and
Nicolas Greiner and Nicolas Chanon for helpful discussions. We are grateful to the Humboldt-Universit\"{a}t zu Berlin
for providing computing resources. This work has been supported by the Alexander von Humboldt Foundation,
in the framework of the Sofja Kovaleskaja Award 2010, endowed by the German Federal Ministry of Education and Research.
}

\providecommand{\href}[2]{#2}\begingroup\raggedright\endgroup

\end{document}